\documentclass[aps,twocolumn,prl,showpacs,floatfix,superscriptaddress]{revtex4}
\usepackage{graphicx}
\usepackage{amsmath}
\usepackage{amsfonts}
\usepackage{amssymb}

\begin{document}

\title{CaMn$_{2}$Sb$_{2}:$ a fully frustrated classical magnetic system}
\author{I. I. Mazin}

\affiliation{Code 6393, Naval Research Laboratory, Washington, DC 20375, USA}

\date{8/9/13}

\begin{abstract}
We show by means of density functional calculations that CaMn$_{2}$Sb$_{2}$
is very close to a mean-field critical point known for the classical
Heisenberg model on the honeycomb lattice. Three entirely different long
range ordered magnetic phases become degenerate at this point: the Neel
phase and two different spiral phases. We speculate that the unusual physical
properties of this compound, observed in recent experiments, in particular
the enigmatic intermediate temperature phase, are due to this
proximity.
\end{abstract}

\maketitle

\affiliation{Code 6393, Naval Research Laboratory, Washington, DC 20375, USA}

\begin{figure}[ptb]
\begin{center}
\includegraphics[width=0.49\columnwidth,angle=0]{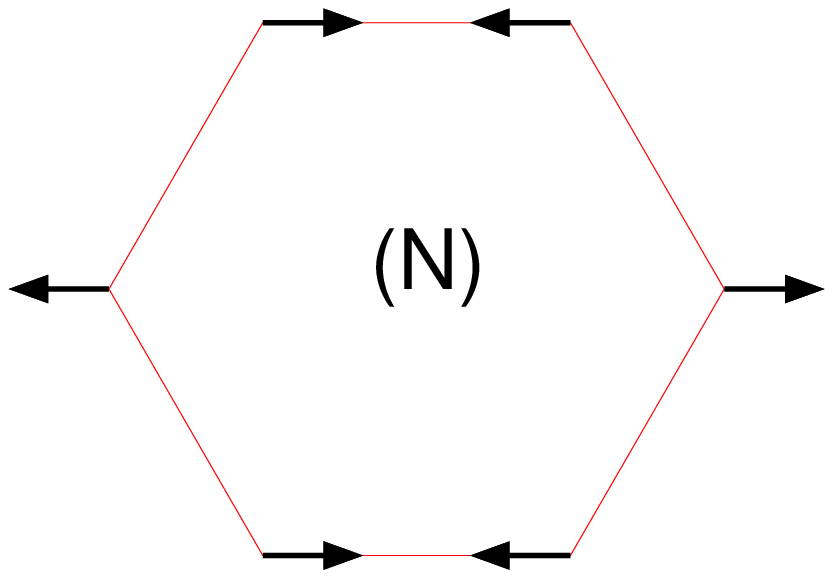}
\includegraphics[width=0.49\columnwidth,angle=0]{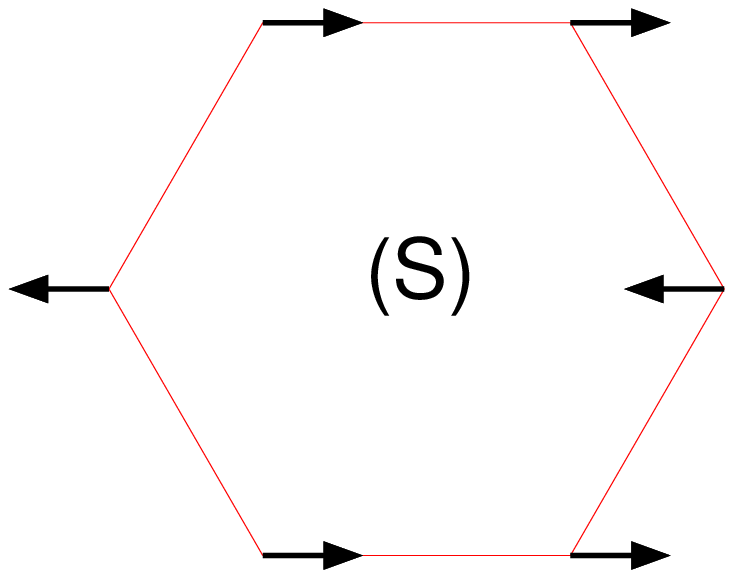}
\includegraphics[width=0.49\columnwidth,angle=0]{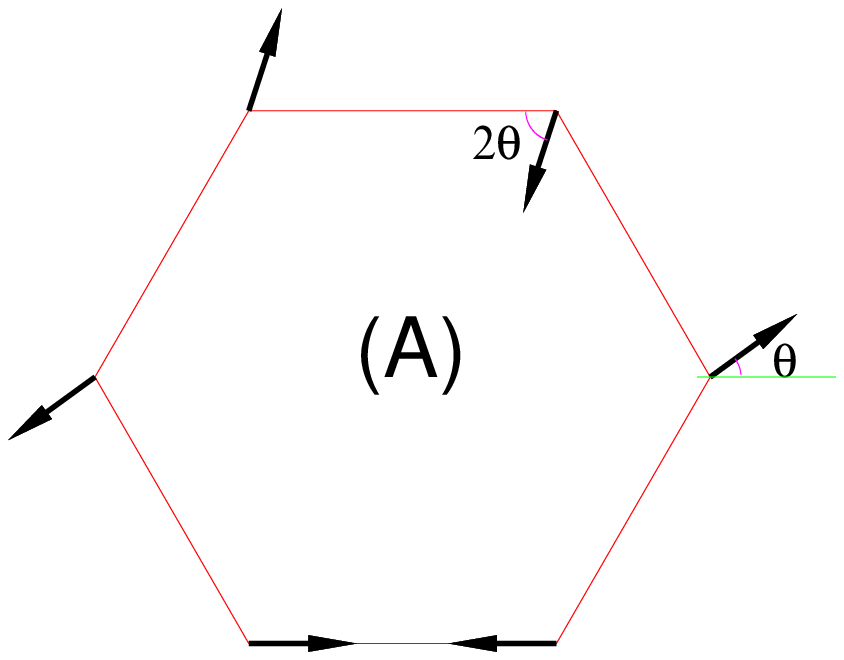}
\includegraphics[width=0.49\columnwidth,angle=0]{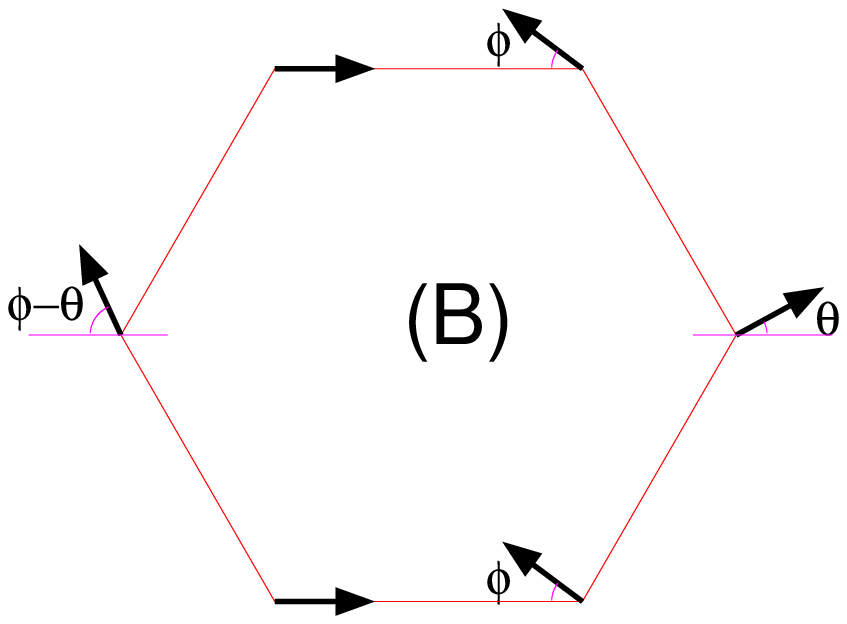}
\end{center}
\caption{ Four antiferromagnetic patterns discussed in the text. Note that 
the spiral A coincides with the phase N at $\theta
=0$ and spiral B with the phase N N  at $\phi=\pi$ and $\theta
=0$, and with S at $\phi=0$ and $\theta=\pi$. At the same time 
the two spirals cannot be continuously transformed into each other
except in this limiting cases.
}%
\label{patterns}%
\end{figure}

\textit{Introduction: }CaMn$_{2}$Sb$_{2}$ is an interesting implementation
of a nearly classical spin system on a honeycomb lattice. Such systems have
been attracting the attention of researchers since the seminal paper of Rastelli 
$et$ $al$ from 1979\cite{Rastelli}. The basic model involves the nearest
neighbor (nn) interaction $J_{1}$ and the second neighbor interaction $J_{2}.$
In the classical limit, the ground state of this $J_{1}-J_{2}$ model, for
antiferromagnetic $J_{1,2}$, is the Neel phase where all nearest neighbor
bonds are fully antiferromagnetic, if $J_{2}/J_{1}<1/6.$ Interestingly, for 
$1/6<J_{2}/J_{1}<1/2,$ there are two degenerate solutions, corresponding to
two different spiral phases (in one of them the two spins in one unit cell
are always antiparallel, and the spiral propagation vector is perpendicular
to a nn bond, in the other the two sublattices are rotated by a particular
angle with respect to each other, and the spiral vector is parallel to a
bond). At $J_{2}/J_{1}>1/2$ the ground state is the stripe phase with
alternating FM pairs (Fig. \ref{patterns}). Both critical points, $%
J_{2}/J_{1}=1/2$ and $J_{2}/J_{1}=1/6$ are triple points: in case there is
an additional parameter, for instance the third neighbor exchange $J_{3},$
three phases meet at these points: the Neel phase and the two spiral phases
at $J_{2}/J_{1}=1/6,$ and the stripe phase and the two spiral phases at $%
J_{2}/J_{1}=1/2.$ Whenever a third parameter is added, be it the third
neighbor exchange\cite{Rastelli}  $J_{3}$, uniaxial
anisotropy\cite{Rastelli} or biquadratic coupling\cite{I} $K,$
complex phase diagrams arise, with the four phases described above, and
additional phases such as zigzag antiferromagnetism.

\begin{figure}[ptb]
\begin{center}
\includegraphics[width=0.59\columnwidth,angle=0]{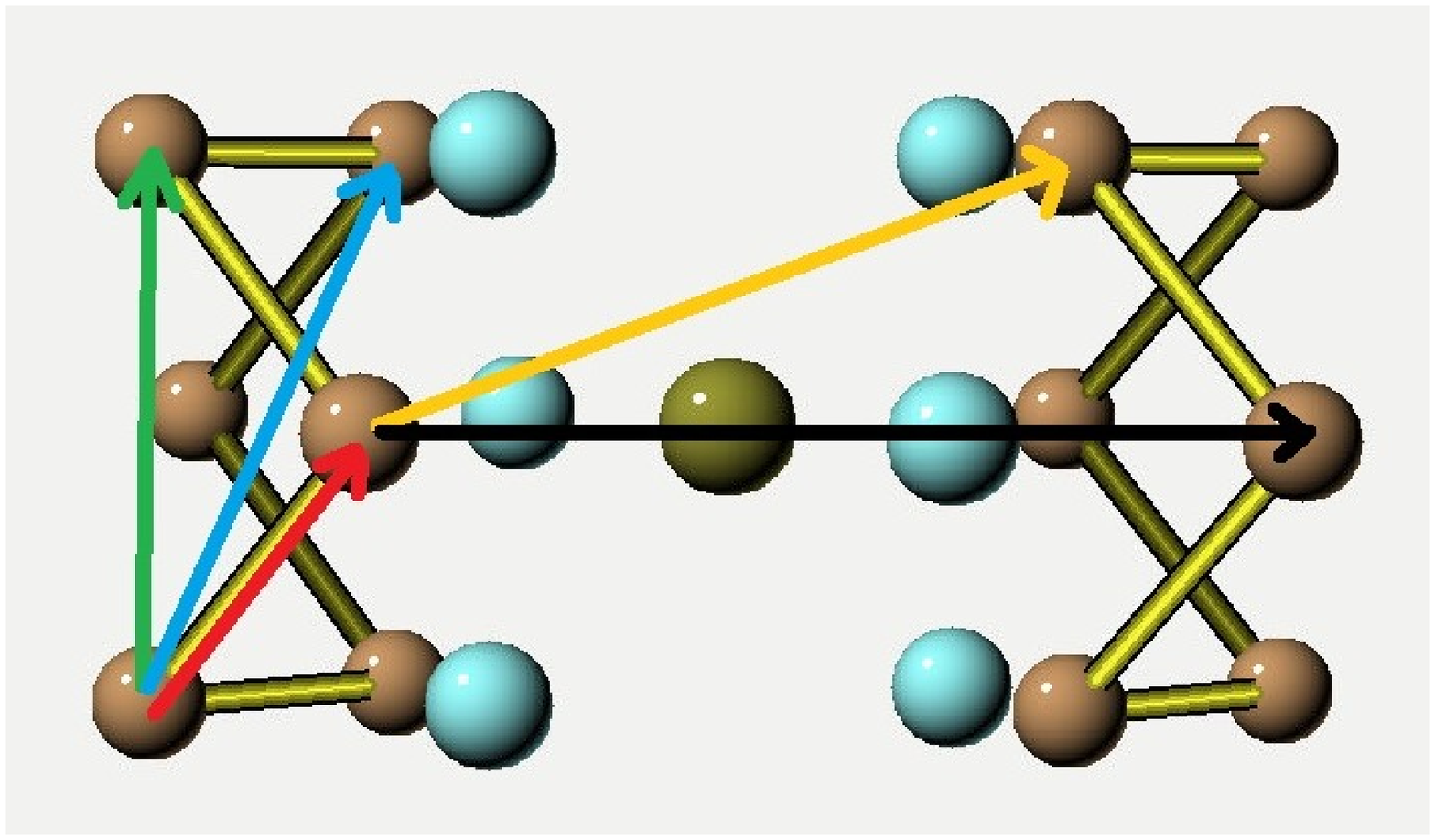}
\includegraphics[width=0.39\columnwidth,angle=0]{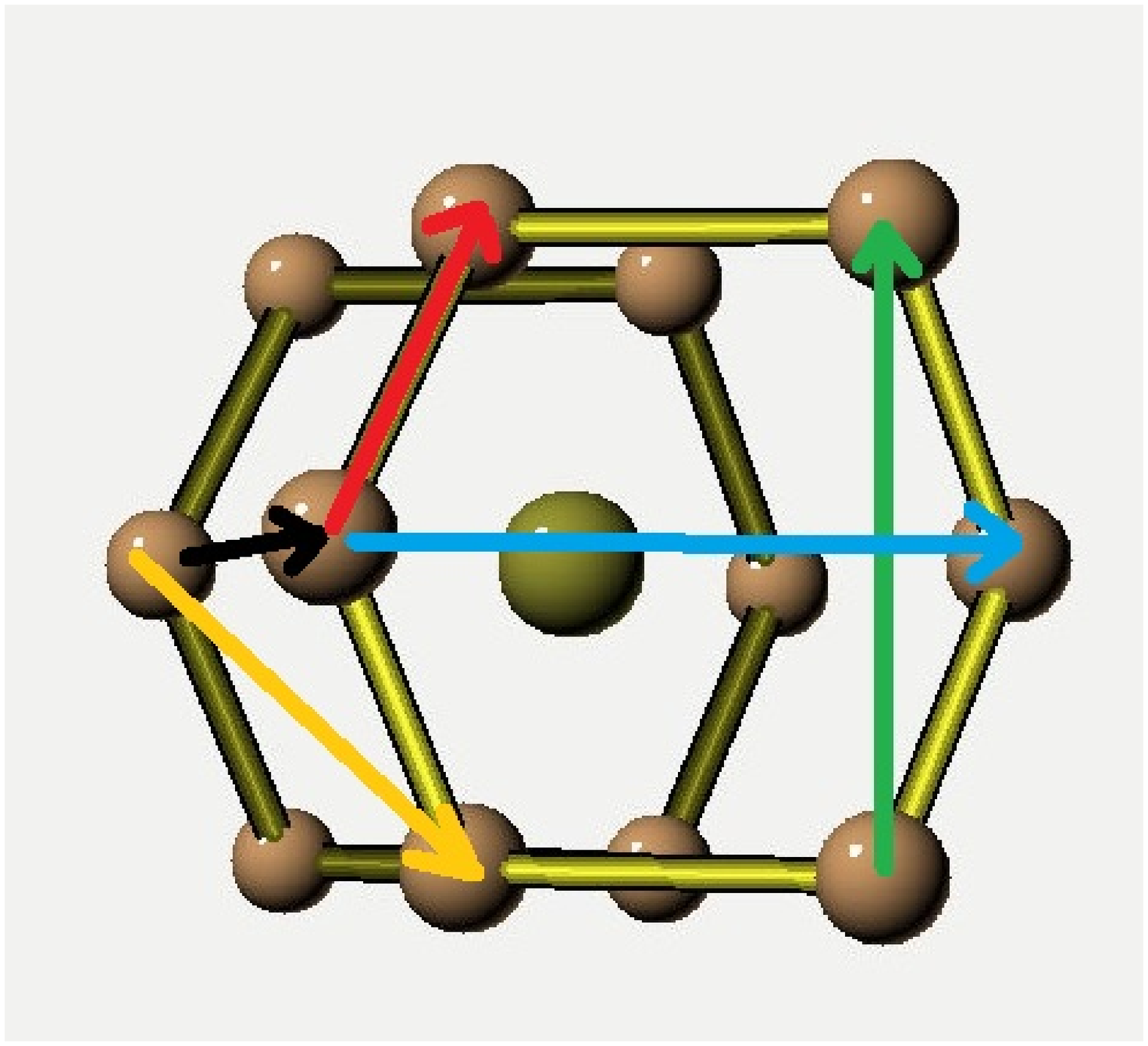}
\end{center}
\caption{(color online) Crystal structure of CaMn$_{2}$Sb$_{2}$, showing examples of the
nearest neighbors ($J_1$, red), 2nd neighbors ($J_2$, green) and 3rd neighbors
($J_3$), blue, inside a MnSb layer, as well as the nearest 
($J_{z1}$, yellow) and 2nd ($J_{z2}$, black) interplanar neighbors. Mn atoms 
are shown by gray spheres, Sb by cyan ones and Ca by green ones.}
\label{structure}%
\end{figure}

The crystal structure of CaMn$_{2}$Sb$_{2}$ is shown in Fig. \ref{structure}%
. The Mn sublattice consists of honeycomb layers in which every other atom is
shifted perpendicular to the plane. This interesting geometry leads to a
rather short Mn-Mn distance, 3.18 \AA , with substantial direct overlap
between the Mn d orbitals and as a result sizable direct antiferromagnetic
exchange. In addition, there are two superexchange paths available. One
connects the nearest neghbor Mns, with an Mn-Sb-Mn angle of 70$^{\circ },$
and the other connects the second neighbors, with an angle of 108$^{\circ }.$
Mn in this compound has valency 2+, and therefore in the high spin state it
would have a magnetic moment of 5 $\mu _{B},$ reduced by hybridization and
fluctuations. Experimentally at low temperature the $J_{1}$-driven Neel
phase was found, with a magnetic moment of 2.8-3.4 $\mu _{B}$. At $T=85$ K
this phase gives rise to another phase of unknown origin, which
experimentally resembles weak ferromagnetism, but no long range
antiferromagnetic order was detected. Finally, at $T\gtrsim 200$ K the
material becomes paramagnetic, exhibiting at $T\gtrsim 300$ K a Curie-Weiss
behavior with an effective moment $M_{CW}=1.4$ $\mu _{B}$. The unusual
character of the intermediate temperature phase, as well as the very low
Curie-Weiss moment, suggest that frustration characteristic of honeycomb
magnetic models may play a role. The low-temperature phase shows an
activation transport behavior with an activation gap $\sim 40$ meV, while
the intermediate temperature phase exhibits a strong increase (up to a factor of 100) of
the resistivity.\cite%
{Aronson} 

We have performed first principle calculations of the electronic and
magnetic properties of CaMn$_{2}$Sb$_{2}$. We found Mn to be in the high
spin state and $d({5\uparrow })$ configuration. We also found a small gap
consistent with the experiment, and magnetic interactions dominated by the
nearest neighbor exchange. Most interestingly, we found that the second
neighbor exchange is 4-6 times smaller than the nearest neighbor one, while
the third neighbor exchange and the biquadratic exchange are very small, and
the magnetic anisotropy is of the easy-plane type. In this regime classical
spins on the honeycomb lattice are highly frustrated, with two spiral and two
collinear phases nearly degenerate. We speculate that this frustration is
the cause of the unusual magnetic phase diagram.

\textit{Calculations.} The calculations in this paper were performed with
the Linear Augmented Plane Wave code WIEN2k\cite{WIEN2K}, using the
following crystallography: symmetry group \#164, $P\bar{3}m1,$ $a=$4.522 \AA %
, $c=$7.458 \AA , $z_{Ca}=0,$ $z_{Mn}=0.3784,$ $z_{Sb}=0.7487,$ and a
Generalized Gradient Approximation for the exchange-correlation potential.
In agreement with previous calculations\cite{Aronson} we found that Mn in
strongly polarized, and Mn moments are well localized, as demonstrated
by the fact that all magnetic configurations converge to about the same
magnetic moment, $\sim 4$ $\mu _{B},$ and the calculated exchange energies
are much smaller than the magnetization energy. The lowest energy among
various collinear states has the Neel state, in which the nn Mn have
opposite spins. 

\begin{figure}[ptb]
\begin{center}
\includegraphics[width=0.99\columnwidth,angle=0]{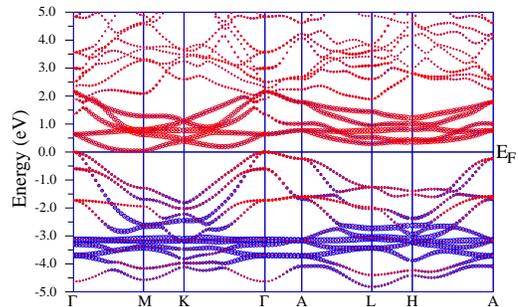}
\end{center}
\caption{(color online) Calculated band structure of CaMn$_{2}$Sb$_{2}$ in the 
antiferromagnetic Neel phase. The symbol size is proportional to the Mn character
(spin-up in blue, below the Fermi level, and spin-down in red, above the Fermi level.   
}%
\label{bands}%
\end{figure}

\begin{figure}[ptb]
\begin{center}
\includegraphics[width=0.99\columnwidth,angle=0]{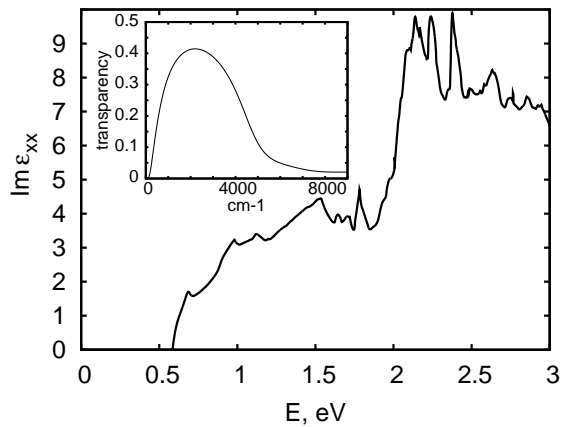}
\end{center}
\caption{The calculated dielectric function of CaMn$_{2}$Sb$_{2}$.
The inset shows the corresponding transmission coefficient (the thickness of the slab
was chosen to roughly match the amplitude of the transmission coefficient in Ref. 
\protect\cite{Aronson}i, and a uniform week absorption was added to account 
for the experimentally observed in-gap states).
}%
\label{opt}%
\end{figure}

The band structure and density of states in this configuration are shown in
Fig. \ref{bands}. One can clearly see that all five $d$ orbitals of Mn are
fully polarized, while the reduction of the moment from 5 to 4 $\mu _{B}$ is
due to hybridization. There is also an indirect excitation gap (of the
spin-flip nature), about 50 meV, consistent with the experimentally measured one.
The calculated optical gap (minimal direct transition energy) is about 0.7
meV, but the absorption at this energy starts relatively slow (Fig. \ref{opt}%
), owing to the fact that the top of the valence band and the bottom of the
conductivity band are formed in the different spin channels (in a
ferromagnetic case, these transtions would have been optically forbidden; in
an antiferromagnetic one, they are allowed but weak). The main absorption edge
is located at 2 eV. Compared to experiment\cite{Aronson}, the calculated
transmission drops at a lower frequency. However, the optical experiments
reported in Ref. \cite{Aronson} should be considered as preliminary. One has
to wait for more state-of-the-art optical data, preferably for the
conductivity, for a full comparison.

In order to understand better magnetic interactions in CaMn$_{2}$Sb$_{2},$
we have calculated the total energies of 10 different magnetic
configurations, and mapped the results onto the model Heisenberg that
includes 1st, 2nd and 3rd nearest neighbors inside the MnSb layers ($J_{1},$ 
$J_{2},$ and $J_{3})$, and 1st and 2nd neighbors between the layers ($J_{z1}$
and $J_{z2}).$ The corresponding distances and values are listed in the
Table. Note that the definition of $J$ in this paper is such that the pair
interaction is $J\sigma _{i}\sigma _{j},$ where $\sigma =\pm 1.$

\begin{table}[ptb]
\caption{Calculated exchange parameters and the corresponding Mn-Mn bond lengths.
See Fig.\protect\ref{structure} for graphical examples.
}
\label{Table}
\begin{center}%
\begin{tabular}{l|lllll}
\hline\hline
& $J_{1}$ & $J_{2}$ & $J_{3}$ & $J_{z1}$ & $J_{z2}$ \\ 
$d$ (\AA ) & 3.179 & 4.522 & 5.528 & 6.219 & 7.458 \\ 
$J$ (meV) & 91 & 22 & 5 & 11 & 3%
\end{tabular}
\end{center}
\end{table}

Since the number of energy differences (9) is larger than the number of
fitted exchange constants, one can estimate the accuracy of the fitting from
the standard fit error. This is on the order of 7--8 meV. Note that on the
mean field level $J_{z}$s simply renormalize the planar interactions, in
other words, assuming perfect interplanar ordering, one can map the 3D
problem onto 2D with $J_{1eff}=J_{1}+J_{z1}=102$ meV. 

We have also looked for other possible deviations from the Heisenberg model:
the biquadratic interaction, which plays an extraordinary important role in
Fe-based superconductors, and single-site anisotropy. The former appears to
be zero within computational accuracy, in other words, for all possible
angles between the Mn spins in the same unit cell the energy is perfectly
cosinusoidal, $E=const+0.5J_{1}\cos \alpha .$ The latter (magnetic anisotropy)
is zero in plane
(although formally hexagonal symmetry allows for magnetic anisotropy when the
field is rotated by 30$^{o},$ for all practical purposes this effect is
negligible), and about 2--3 meV/Fe otherwise, with the direction
perpendicular to the plane being the hard axis.

\textit{Discussion}. The fact that essentially any magnetic pattern can be
stabilized in the calculations, and that the energy cost of suppressing
magnetism entirely (0.8 eV/Fe) is much larger than the exchange constants
indicates that  CaMn$_{2}$Sb$_{2}$ should be considered a local moment
system, and overall one expects Mn to be subject to considerable Hubbard
correlations. The fact that the nearest-neighbor exchange follows the
Heisenberg formula nearly exactly also speaks in favor of classical
superexchange and the Hubbard model. Yet DFT calculations reproduce the
excitation gap well, in fact, better than dynamic mean field calculations%
\cite{Haule}. The reasons for such an unexpected success of the DFT are unclear
st the moment. Given the sharp increase of the resistivity in the other, intermediate temperature
phase, one may think that the Neel phase is affected by some cancellation of errors,
not operative in the other phases, which leads to an effective increase of the 
excitation gap in the intermediate phase.

\begin{figure}[ptb]
\begin{center}
\includegraphics[width=0.99\columnwidth,angle=0]{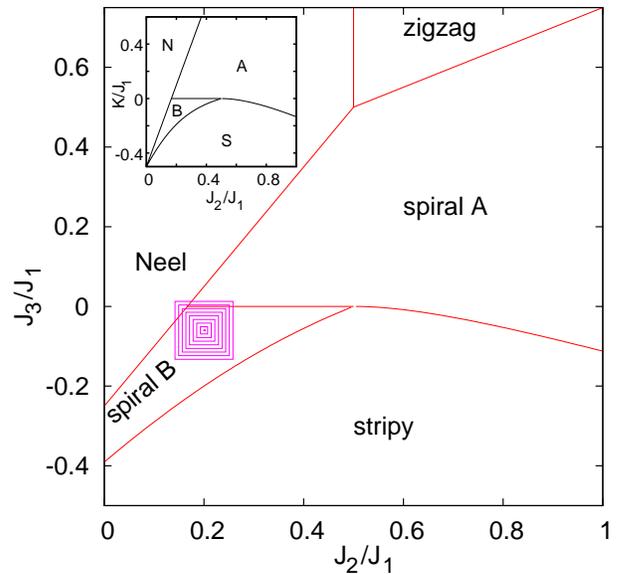}
\end{center}
\caption{Phase diagram of the classical $J_1--J_2--J_3$ Heisenberg model. The phase boundaries 
are given by the lines $
y=(x-\sqrt{x^2+2(x-1/2)^2})/2$, $
y=(1/2+3x-\sqrt{51/12-5x+9x^2})/4$, $
y=(6x-1)/4$, $
y=(2x+1)/4$, $
y=0$, $
x=1/2$\protect\cite{Rastelli}. The square shows the region in the phase diagram where, within computational 
accuracy, the actual system is located. The insert shows the phase diagram for the 
$J_1-J_2-K$ model, where K is the biquadratic coupling. The phase boundaries are:
$y=(3x-\sqrt{1-2x+9x^2})/2$, $
y=\sqrt{x-1/4}-x$, $
y=(6x-1)/2$, $
y=0$\protect\cite{I}.
}%
\label{PhD}%
\end{figure}

In the classical phase diagram of an antiferromagnetic $J_{1},J_{2}$ 2D
honeycomb layer there are five planar phases: The Neel (N) phase, where all
nearest neighbors order antiferromagnetically, the \textquotedblleft
stripy\textquotedblright\ (S) phase where all bond along one direction are
ferromagnetic, and forming double stripes order antiferromagnetically,
and two spiral phases (A and B) described in the introduction. For $%
J_{2}/J_{1}<1/6$ the ground state is N, for $1/6<J_{2}/J_{1}<1$ the ground
state is degenerate between phases A and B, and for the larger $J_{2}$ the
ground state is S.  Adding $J_{3}$ creates a complicated phase diagram, with
a new \textquotedblleft zigzag\textquotedblright\ phase emerging at large $%
J_{3},$ which is shown in Fig. \ref{PhD}, where we also for completeness show the phase
diagram $J_{1},J_{2},K$ with a biquadratic interaction, not published before.
In this phase diagram,
the calculated interactions correspond to the point $J_{2}/J_{1eff}\sim 0.21$%
, $J_{3}/J_{1eff}\sim 0.05,$ but the error in these numbers is close to $\pm
0.08.$ In other words, according to the calculations the material is
extremely close to a highly frustrated critical point where three different
phases are degenerate, the Neel phase and two qualitatively different spiral
phases, $J_{2}/J_{1}=1/6,$ $J_{3}=0.$ 

The thermodynamics of the classical Heisenberg model on the honeycomb lattice
has not, to our knowledge, been systematically studied and this subject is
beyond the scope of the current paper. It is however likely that the phase
boundaries between the zero-temperature phase shift as the entropy is
included, so it is tempting to ascribe the transformation at 85 K to a phase
transition between the Neel and one of the two spiral phases (more likely the
spiral A). Very weak net ferromagnetism is then due to relativistic effects,
which can be estimated by the ratio between the magnetic anisotropy energy
(2-3 meV) and magnetization energy (800 meV), which is $\sim 3\times 10^{-3},
$ to be compared with the experimental ferromagnetic moment of $7\times
10^{-3}$ $\mu _{B}\approx 3\times 10^{-3}$ $M_{Mn}.$ 

No long range order has been observed between 85 and 210 K in neutron
scattering; this suggests that the spirals break after relatively short
distance and the emerging spiral chunks are randomly distributed over the
three equivalent crystallographic directions. Such a state would be very
similar to the \textquotedblleft magnetic liquid\textquotedblright\ state in
MnSi\cite{Reznik}, also called a \textquotedblleft
cholesteric\textquotedblright\ phase, or a \textquotedblleft
blue\textquotedblright\ phase in the original paper. This state is observed
in MnSi near the phase transition into a long-range ordered spiral phase, on
the high-temperature (high-pressure) side of the phase transition. Neutron
scattering in this phase reveals no long range order, but well-defined
spirals of considerable length (several hundred \AA ), propagating in all
crystallographically equivalent directions with equal weight. It is still
not clear whether the spirals in the \textquotedblleft blue
phase\textquotedblright\ of MnSi form domains (possibly dynamic) or are
meandering around, periodically switching directions. Both options are
open as well in the intermediate temperature phase of CaMn$_{2}$Sb$_{2}.$

Another corollary of proximity to the critical point at $J_{2}/J_{1}=1/6$ is
that unusual low energy magnetic excitations should be present in the
low-temperature Neel phase. More detailed experimental spectroscopic studies
are highly desirable.

To summarize, we have shown that CaMn$_{2}$Sb$_{2}$ is extremely close
(within computational accuracy) to a critical point where three
entirely different magnetic phases are degenerate on the mean field level.
We suggest that unusual properties of this compound, obtained in recent
experiments, are related to this unique proximity.

The author acknowledges discussions with Girsh Blumberg and Meigan Aronson,
and funding from the Office of Naval Research (ONR)
through the Naval Research Laboratory's Basic Research Program, and from the
Alexander von Humboldt Foundation.

\end{document}